\documentclass[journal]{IEEEtran}

\ifCLASSINFOpdf

\else

\fi

\usepackage{booktabs}
\usepackage{graphicx}
\usepackage{enumerate}
\usepackage{subfigure}
\usepackage{cite}

\usepackage{algorithm}
\usepackage{algorithmic}
\usepackage{graphics}

\usepackage{indentfirst}
\usepackage{verbatim}

\usepackage{amssymb}
\usepackage{longtable}
\usepackage{epsfig} %% for loading postscript figures
\usepackage{amsthm}
\usepackage{mathrsfs}
\usepackage{amsmath}
\usepackage{stfloats}
\usepackage{multirow}
\usepackage{caption}
\usepackage[hang,flushmargin]{footmisc}

\newtheorem{assumption}{Assumption}

\begin{document}

	\bibliographystyle{IEEEtran}

	\title{An attention-based unsupervised adversarial model\\ for movie review spam detection}

\author{Yuan~Gao,
	Maoguo~Gong,~\IEEEmembership{Senior Member,~IEEE,}
	Yu~Xie,
	and~A.~K.~Qin,~\IEEEmembership{Senior Member,~IEEE}% <-this % stops a space
	\thanks{
		Y. Gao, M. Gong and Y. Xie are with the School of Electronic Engineering, Key Laboratory of Intelligent Perception and Image Understanding of Ministry of Education, Xidian University, Xi'an, Shanxi Province 710071, China. (e-mail: cn\_gaoyuan@foxmail.com; gong@ieee.org; sxlljcxy@gmail.com)

		A. K. Qin is with the department of Computer Science and Software Engineering, Swinburne University of Technology, Melbourne, Australia. (e-mail: kqin@swin.edu.au)}% <-this % stops a space
}
	

	\markboth{IEEE TRANSACTIONS ON MULTIMEDIA}%
	{Shell \MakeLowercase{\textit{et al.}}: Bare Demo of IEEEtran.cls for IEEE Journals}



	\maketitle
	
\begin{abstract}
With the prevalence of the Internet, online reviews have become a valuable information resource for people. However, the authenticity of online reviews remains a concern, and deceptive reviews have become one of the most urgent network security problems to be solved. Review spams will mislead users into making suboptimal choices and inflict their trust in online reviews. Most existing research manually extracted features and labeled training samples, which are usually complicated and time-consuming. This paper focuses primarily on a neglected emerging domain - movie review, and develops a novel unsupervised spam detection model with an attention mechanism. By extracting the statistical features of reviews, it is revealed that users will express their sentiments on different aspects of movies in reviews. An attention mechanism is introduced in the review embedding, and the conditional generative adversarial network is exploited to learn users' review style for different genres of movies. The proposed model is evaluated on movie reviews crawled from Douban, a Chinese online community where people could express their feelings about movies. The experimental results demonstrate the superior performance of the proposed approach.
\end{abstract}


% \fi
\IEEEpeerreviewmaketitle

\section{Introduction}
\IEEEPARstart{W}{ith} the emerging and developing of Web 2.0 that emphasizes the participation of users, consumers increasingly depend on user-generated online reviews to make purchase or investment decisions \cite{chua2016helpfulness} \cite{zhao2018exploring} \cite{li2019effect}. Online reviews enable consumers to post reviews describing their experiences with products and improve the ability of users to evaluate unobservable product quality. However, these growing text information is also full of various uncontrollable risk factors that needs to be seriously treated and managed by websites and platforms. A significant impediment to the usefulness of reviews is the existence of review spam, which is designed to give an unfair view of products and influence users' perception of them by inflating or damaging their reputation \cite{hu2011manipulation} \cite{hu2011fraud}. Both over-rating and under-rating affect the sales performance of products, and it is thus an important task to detect review spam to protect the genuine interests of consumers and product vendors.

In the past few years, there has been a growing interest in review spam detection from both academia and industry \cite{dewang2018state} \cite{xue2019content}. A preliminary investigation is given in \cite{jindal2008opinion}, where three types of review spam are identified, including untruthful reviews (reviews that maliciously promote or defame products), reviews on brands but not products, and non-reviews (e.g. advertisements). Since they have no manually labeled training instances, untruthful review detection is performed by using duplicate reviews as labeled deceptive data, and a regression model is built on these samples. Most subsequent methods utilize supervised or semi-supervised models under its influence by manually identifying review spam and extracting features \cite{mayzlin2014promotional} \cite{saidani2017supervised} \cite{khurshid2017recital}. However, it is incredibly time-consuming and impractical for service providers to label previous spammers, because a spammer will carefully craft a review that is just like other innocent reviews \cite{ott2011finding}, and even a small number of incorrect annotations may bring poor results of the model. Moreover, feature identification and construction is also tricky and complicated. Therefore, it is necessary to propose an unsupervised way for review spam detection.

Most existing works on review spam detection focus on product reviews. Compared with online product review platforms, online movie review platforms appear later and thus receive little attention. This paper focuses primarily on movie review spam detection. Top reviews and online word-of-mouth have a significant impact on the box office performance of movies \cite{ma2019analyzing} \cite{legoux2016effect}, while the censorship is not strict and anyone can post reviews and express their sentiments online. Under the substantial commercial interest, more and more review spammers make deceptive critics to increase the evaluation of the target movie deliberately and discredit the competition movies arbitrarily, which will mislead users who plan to watch movies and attract publishers to put funds into review manipulation rather than filmmaking. For these reasons, detecting movie review spams is essential and urgent for the movie industry. Different from product reviews, movie reviews have the following characteristic. Due to the subjectivity of movie reviews, it is a challenging task to label them manually and therefore supervised methods are no longer applicable. Furthermore, carefully crafted deceptive reviews usually look perfectly normal, thus the artificially extracted features, such as text similarity or length of review, are ineffective. Besides, movie reviews usually entail unique movie plots or actors, which increase the burden of spam detection. As a result, spam detection of movie reviews is more challenging than that of product reviews.

To address the aforementioned challenges, the paper proposes a novel movie review spam detection method using an attention-based unsupervised model with conditional generative adversarial networks. Unlike the previous approaches for review spam detection, the presented user-centric model can achieve outstanding performance under the circumstance of movie reviews. The characteristics of movie reviews are discussed in depth, which naturally raise several assumptions about deceptive reviews. Meanwhile, it is verified that the user's attention (e.g. plot, actors, visual effects) are always distinct for different genres of movies. During the process of mapping reviews into low-dimensional dense vectors, the weights of feature words will increase, and what users are interested in about a movie can be captured. Afterward, the attention-driven conditional generative adversarial network (adCGAN) is utilized to identify the specific condition-sample pairs with movie elements (e.g. score, genres, region) being conditions and review vectors being samples. Particularly, the generator aims to misjudge the discriminator and generate vectors that have similar distributions as real reviews with certain conditions. The discriminator judges whether movie-review pairs are real, where conditions have been resampled, thus correlations are extracted and deceptive reviews will be identified. To the best of our knowledge, this paper is the first to explore the potential of GANs for review spam detection. Specifically, the main contributions of this paper are as follows:
\begin{itemize}
\item{The paper investigates the characteristics of movie review spams and the differences between product and movie review spams, then several novel techniques for deceptive review detection are presented. Compared with traditional review spam detection algorithms, the approach can effectively detect deceptive opinion spam under the circumstance of movie reviews.}
\item{The paper introduces an attention mechanism and a user-centric model, which is preferred over the review-centric one as gathering behavioral evidence of users is more effective than features of deceptive reviews. The weights of words change adaptively according to users’ attention towards different aspects of movies in sentence embedding, thus user preferences can be extracted and encoded into the vector representations of reviews.}
\item{It alleviates the problem of lacking labeled deceptive reviews by introducing unmatched movie-review pairs as fake samples, so that the conditional generative adversarial network is able to learn user’s language style from previous reviews and under what conditions will users write certain reviews.}
\item{The paper explores several ways to identify spammed movies and reviews, and our method is evaluated on movie reviews crawled from Douban \footnote{https://movie.douban.com}, the most authoritative online movie rating and review website in China. Experimental results demonstrate that the proposed algorithm outperforms other anomaly detection baselines.}
\end{itemize}

The remainder of this paper is organized as follows. Section~\uppercase\expandafter{\romannumeral2} briefly presents the related backgrounds and algorithms about review spam detection, word embedding and generative adversarial networks. In Section~\uppercase\expandafter{\romannumeral3}, clear and objective definitions of the review spam and the problem of review spam detection are given, then the details of the approach are described. Section~\uppercase\expandafter{\romannumeral4} shows several ways to artificially detect deceptive reviews, as well as extensive experiments to validate the effectiveness. Finally, we conclude with a discussion of the proposed framework and summarize the future work in Section~\uppercase\expandafter{\romannumeral5}.

\section{Related Work}
\label{para:2}
Recently, online review spam detection has been a popular research topic \cite{crawford2015survey} \cite{hussain2019spam} \cite{vidanagama2019deceptive}. Most existing review spam detection techniques are supervised or semi-supervised methods with pre-defined features. As mentioned earlier, Jindal \emph{et al.} \cite{jindal2008opinion} identified three types of spam and applied logistic regression on manually labeled training examples. In particular, they manually selected 35 features, including price and sales rank of the product, percent of positive and negative words in reviews, and the length of review titles, etc. Lim \emph{et al.} \cite{lim2010detecting} investigated several characteristic behaviors of review spammers and modeled these behaviors for spam detection. They also proposed a scoring method to measure the degree of spam for reviewers and applied it on Amazon. Rayana \emph{et al.} \cite{rayana2015collective} proposed a holistic approach that utilized clues from metadata (text, timestamp, rating) and relational data, then harnessed them collectively to spot suspicious users and reviews. Yang \emph{et al.} \cite{YangLCX17} presented a three-phase method to address the problem of identifying spammer groups and individual spammers. They utilized duplicate and near duplicate reviews detection, reviewers interest similarity and individual spammer behavior in spammer detection. Generally, the selected features of models gradually changed from shallow (the length of reviews) to deep (users' interests) with the development of deep learning techniques.

Word embedding, which is a distributed representation method, is widely used in various natural language processing (NLP) tasks, such as opinion analysis \cite{quan2015word} \cite{poria2016aspect} and sentiment classification \cite{tang2014learning} \cite{ji2019cross}. Word embedding methods represent words as continuous dense vectors in a low-dimensional space, which capture lexical and semantic properties of words. Mikolov \emph{et al.} \cite{mikolov2013distributed} proposed SkipGram, in which vectors are obtained from the internal representations from neural network models of text. It optimizes a neighborhood preserving likelihood objective using stochastic gradient descent with negative sampling. To be specific, given a sequence of training words $s = (w_1,w_2,...,w_{|s|})$, the SkipGram model seeks to maximize the average log-probability of observing the context of a word $w_i$ conditioned on its representation $\Phi(w)$. The $\Phi(w)$ is a mapping function that maps a word to low-dimensional dense vector, and its value is the weight from the input layer to the hidden layer of the neural network. The number of neurons in the input layer corresponds to the size of the vocabulary, and that in the hidden layer is the embedding dimension.
\begin{equation}
\max\limits_\Phi\quad\frac{1}{|s|}\sum\limits_{i=1}^{|s|} \sum\limits_{-c\leq j \leq c,j\neq 0} logPr\bigl(w_{i+j}\ |\ \Phi(w_i)\bigr),
\end{equation}
where the context is composed of words on both sides of the given word within a window size $c$. A larger $c$ could result in more training samples and thus lead to higher accuracy, at the expense of the training time. Introducing the randomness, dynamic window size is better than fixed one, as a central word may be related to many or only a few words in a sentence. Since movie reviews are usually not too long in length, the window size is set to a uniform distribution of 1 to 5 in the experiment. Considering the symmetric effect over the current word and context words in the low-dimensional representation space, the conditional probability $Pr\bigl(w_o|\Phi(w_i) \bigr)$ is modeled as a softmax unit:
\begin{equation}
Pr\bigl(w_o\ |\ \Phi(w_i) \bigr)=\frac{exp\bigl(\Phi(w_o)\cdot\Phi(w_i)\bigr)}{\sum_{w\in W} exp\bigl(\Phi(w)\cdot\Phi(w_i)\bigr)},
\end{equation}
where $w_i$ and $w_o$ are the input and output words, and $W$ are words in the vocabulary. The appealing, intuitive interpretation of word vectors enables them to be used in many machine learning algorithms and strategies. 

\begin{figure*}[htb!]
\centering
  \includegraphics[width=0.7\linewidth]{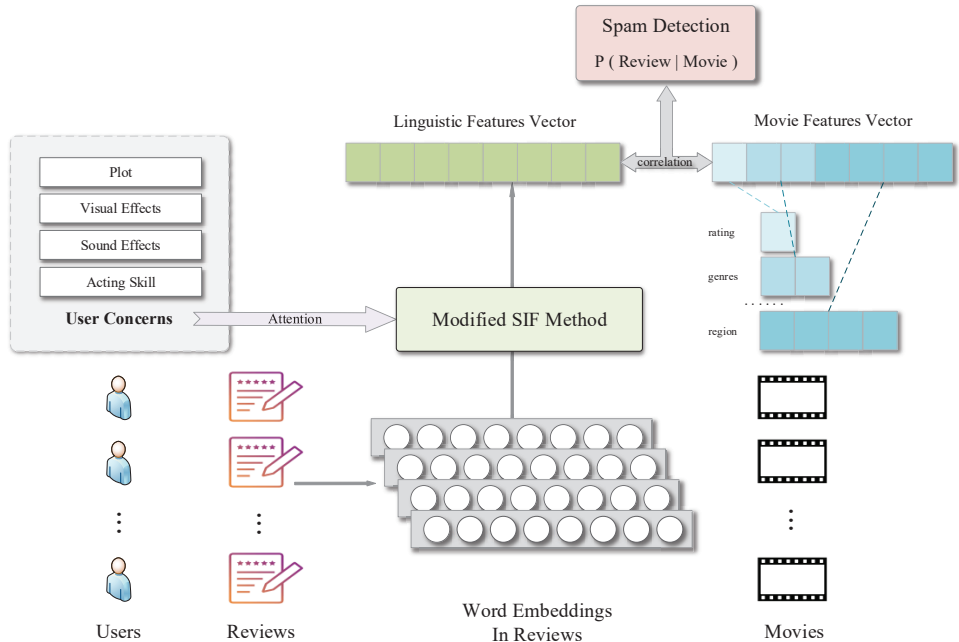}
  \caption{
  A graphical illustration of the review spam detection model. User concerns are embedded into the sentence embeddings leveraging modified SIF method, thus more meaningful linguistic features with attention are extracted. Several critical features of movies, such as user rating, movie genres and region, are encoded and concatenated them as the movie features vector. Afterwards, the matching relation between movies and reviews are learned to discriminate whether a particular movie-review pair is deceptive.}
\label{fig:framework}
\end{figure*}
Inspired by the success of generative adversarial networks \cite{goodfellow2014generative} in the computer vision, recently, the idea of adversarial training has been extensively explored \cite{mirza2014conditional} \cite{arjovsky2017wasserstein}. The adversarially trained model is based on two different networks trained with unsupervised data. One network is the generator ($\mathcal{G}$), which aims at learning the distribution over training data via a mapping of noise samples. The other one is the discriminator ($\mathcal{D}$), which aims at discriminating real data from fake ones generated from $\mathcal{G}$. GANs are capable of learning the distribution of normal data, and then compare the abnormal data with the corresponding generated normal data for anomaly detection. They alleviate the severe problem in anomaly detection of obtaining the ground truth, and enable to detect abnormal samples by computing the difference \cite{ravanbakhsh2017abnormal} \cite{schlegl2017unsupervised}. In this paper, GANs are innovatively applied to review spam detection drawing on the idea of generating reviews and discriminating spams. Specifically, $\mathcal{G}$ is trained to generate embedding vectors of reviews under certain conditions, while $\mathcal{D}$ to learn the distribution of vectors as well as the correlations between movie elements and reviews.

\section{Methodology}
\label{para:3}
In this section, details of the proposed model are presented. The overall architecture of the proposed model is presented in Fig.~\ref{fig:framework}. At first, the SkipGram model is leveraged to learn low-dimensional dense word embeddings from all reviews, and the SIF method calculates the weighted average of word vectors in the sentence and removing the common parts of the corpus to generate sentence embeddings. In the process, user concerns are introduced and the weights of words are further adjusted according to their degree of affiliation with the concerns. The encoded movie features and sentence vectors form the review pairs together, and the conditional generative adversarial network could learn the matching relation and identify the deceptive reviews.

It begins by giving a formal definition of the review spam and the problem of review spam detection, and the inherent correlations between the user's attention and features of movies are demonstrated. After that, it introduces the sentence vector with an attention mechanism, which has a superior performance on similarity measurement. Finally, an attention-driven conditional generative adversarial network framework is proposed, which simultaneously learns to perform review generation and attention matching.

\subsection{Problem Statement}
Review spam detection aims to train an unsupervised model which is able to detect deceptive reviews on a dataset that is highly biased towards a particular class, i.e., comprising a majority of normal non-spam reviews and only a few abnormal reviews. Let $\mathbb{D}$ be the set of unlabeled reviews of a given user comprising $M$ reviews, $\mathbb{D}=\{r_1,r_2,...,r_M\}$, and additional movie information $\mathbb{C}=\{c_1,c_2,...,c_M\}$. Each review $r=\{w_1,w_2,...,w_{|r|}\}$ consists of a sequence of word tokens, where $w_r \in R$ reprensets the $r^{th}$ token in the sentence and $R$ is a corpus of tokens. Review spam detection attempts to identify whether a review about a movie in $\mathbb{D}$ is a spam or non-spam, given the auxiliary information (score, genres, region) of this movie.

\begin{figure*}[htb]
  \centering
  \subfigure[$user_1$]{
    \includegraphics[width=0.4\linewidth]{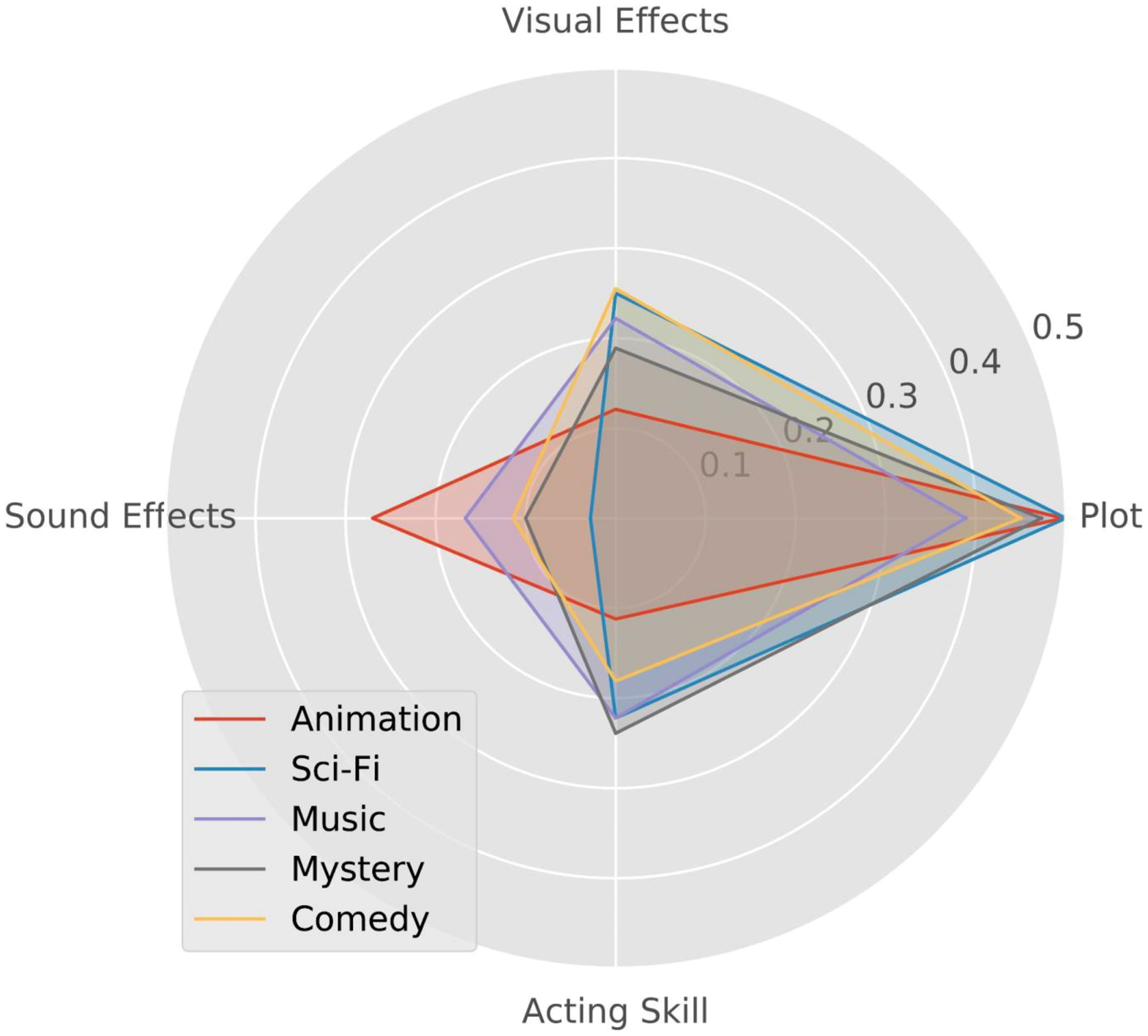}}
  \subfigure[$user_2$]{
    \includegraphics[width=0.4\linewidth]{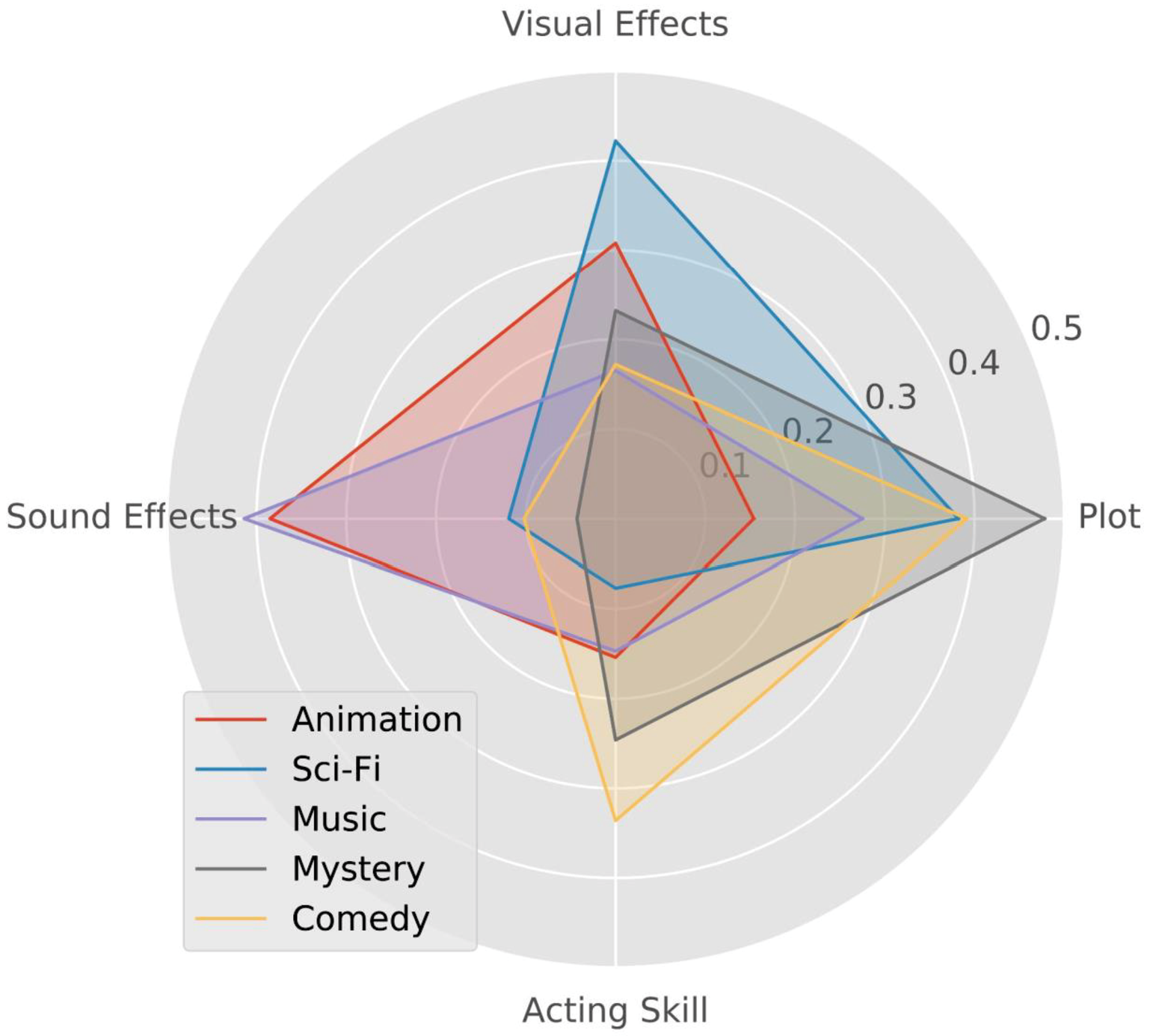}}
  \caption{Users' attention toward four aspects of different genres of movies.}
  \label{fig:attention}
\end{figure*}
Recent researches conclude that online reviews are valuable marketing resources for users and vendors with critical implications for a product's success \cite{yang2016effect} \cite{babic2016effect}. In order to better distinguish between truthful and deceptive reviews, several implicit but essential assumptions are proposed in this paper, which are:
\begin{assumption}
Users' historical reviews are normal and truthful, while their reviews of movies released in recent years are likely to be deceptive reviews \cite{lee2018sentiment} \cite{ulker2018marketing}.
\end{assumption}
Spammers will post deceptive reviews to increase the evaluation of the target movie deliberately, and they may discredit the competition movies arbitrarily. Considering the goal of spammers is to inflate or damage the target movie's reputation so as to affect its box office, there is no need for them to post deceptive reviews to a movie which has been taken out of theaters or released decades ago. Moreover, spammed movies only make up a small percentage of the total number of user reviews, and a few misjudgement will not have much effect on the model since the approach is attention-driven and the credibility is given by conditional probability. Different from the linear model, generative adversarial networks are not sensitive to outliers, and a small number of deceptive reviews does not change the confidence boundary \cite{fawzi2017robustness}. The historical deceptive review will not introduce any error into the results, unless it happens to have the same condition as that of reviews in the test set. Therefore, historical reviews of a user can be utilized as truthful samples to learn the confidence boundary, and those posted in recent years serve as test samples to be detected.
\begin{assumption}
A normal user usually has a relatively consistent attitude toward specific actors and directors, which means a user could be a spammer if he or she gives completely opposite reviews to the same actor or director \cite{zhuang2006movie} \cite{diao2014jointly}.
\end{assumption}
Whether users like or hate an actor, their attitudes are often hard to change. Admittedly, an actor's acting skill may improve or decrease, but it is a gradual process and is therefore acceptable to our model. However, a sudden opposite attitude towards a certain actor or director could be considered deceptive. For example, a user claims that ``He is my favorite actor and I would love to see the premiere for him'', but while checking the user's historical reviews over, it is found out that he only gives moderate reviews to the actor, and there is no doubt that this user is a spammer. On the contrary, fans of an actor may always give praise for his or her movies, regardless of the actual quality. Note that it does not mean that each opposite rating is necessarily a deceptive review. The conditional GAN is able to learn the matching relation to reviews from the input movie features, and automatically judge the potential importance of each feature. Hence, the extracted rating features may not be the deciding factors.
\begin{assumption}
Users will post review spam driven by benifits instead of real feelings, thus it is difficult for them to evaluate the movie from the perspective of normal reviews, and their language style or attention could be distinct from those of previous reviews of similar movies \cite{shehnepoor2017netspam} \cite{li2017document}.
\end{assumption}
The common underlying assumption for semantic-based review spam detection methods is that the language style of deceptive reviews is different from normal ones. For movie reviews, the attention is further incorporated into the linguistic features. Because reviews always embody users' personal tendency of the given movie \cite{sun2016research}, user's sentimental polarities are able to be found by methods of sentiment analysis. Inspired by previous researches \cite{thet2010aspect}, user's attention to a movie is further elaborated into the following four aspects: plot, visual effects, sound effects and acting skill. It is valuable for exploring attractive (and undesirable) aspects, which could help to extract additional information from reviews. In order to validate the effectiveness of the attention mechanism, the method without it, i.e. adCGAN\underline{~~}w/o\underline{~~}attention, has been proposed and compared with our model in Section~\uppercase\expandafter{\romannumeral4}. The method does not embed user’s attention into sentence vectors, and the conditional GAN can only utilize the matching relation between movies and reviews for spam detection.

Fig.~\ref{fig:attention} reveals that users' attention to the four aspects varies according to the genres of movies, and different users tend to have distinct attention to these elements. For instance, $user_2$ is usually interested in the plot of a movie. However, he might be suspected of being a spammer if he suddenly focuses on praising or criticizing the sound effects of a Sci-Fi movie. Besides, the user profile will be clearer and more authentic by combining other features.

\begin{figure*}[htb]
  \centering
  \subfigure[$user_1$]{
    \includegraphics[width=0.35\linewidth]{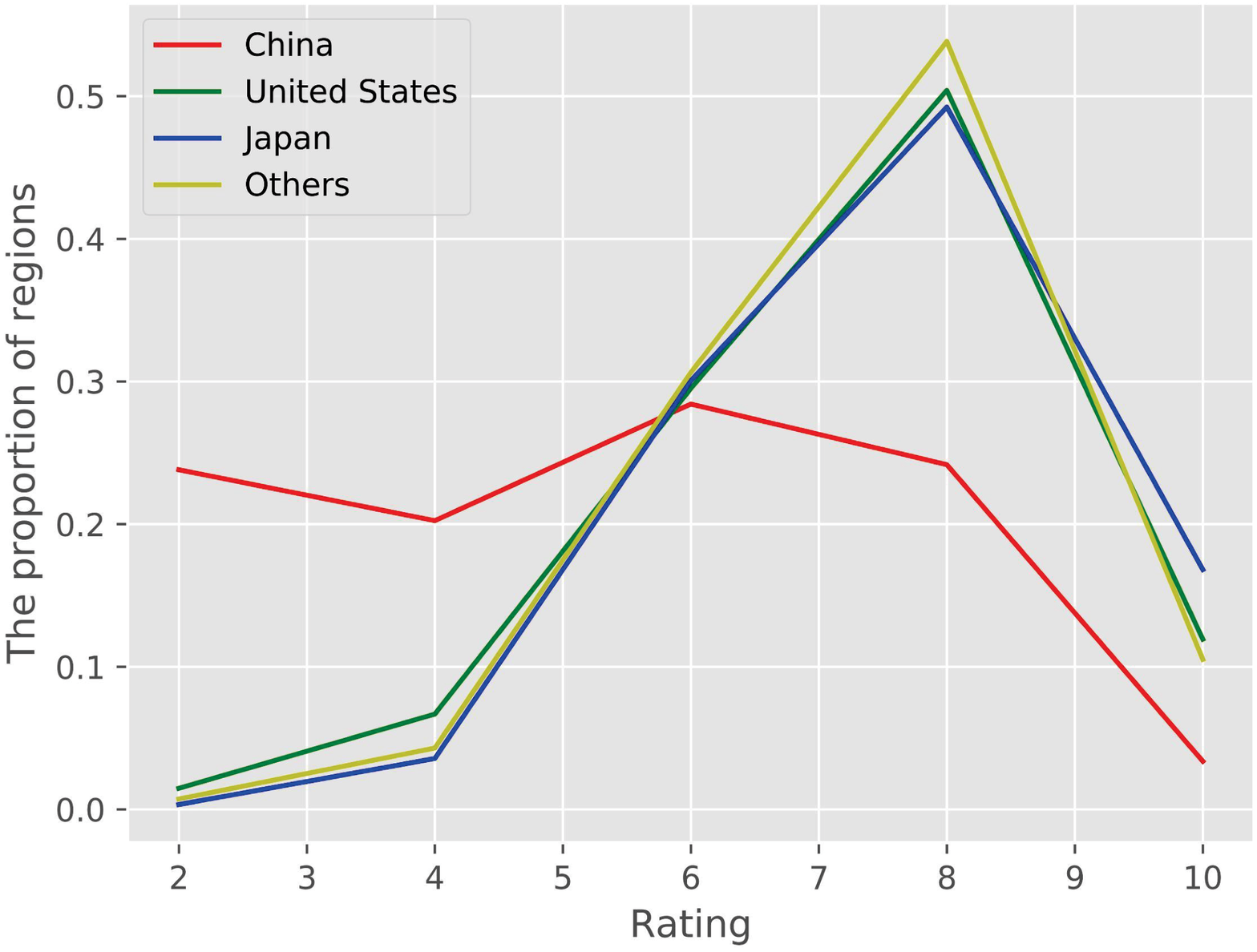}}
  \subfigure[$user_2$]{
    \includegraphics[width=0.35\linewidth]{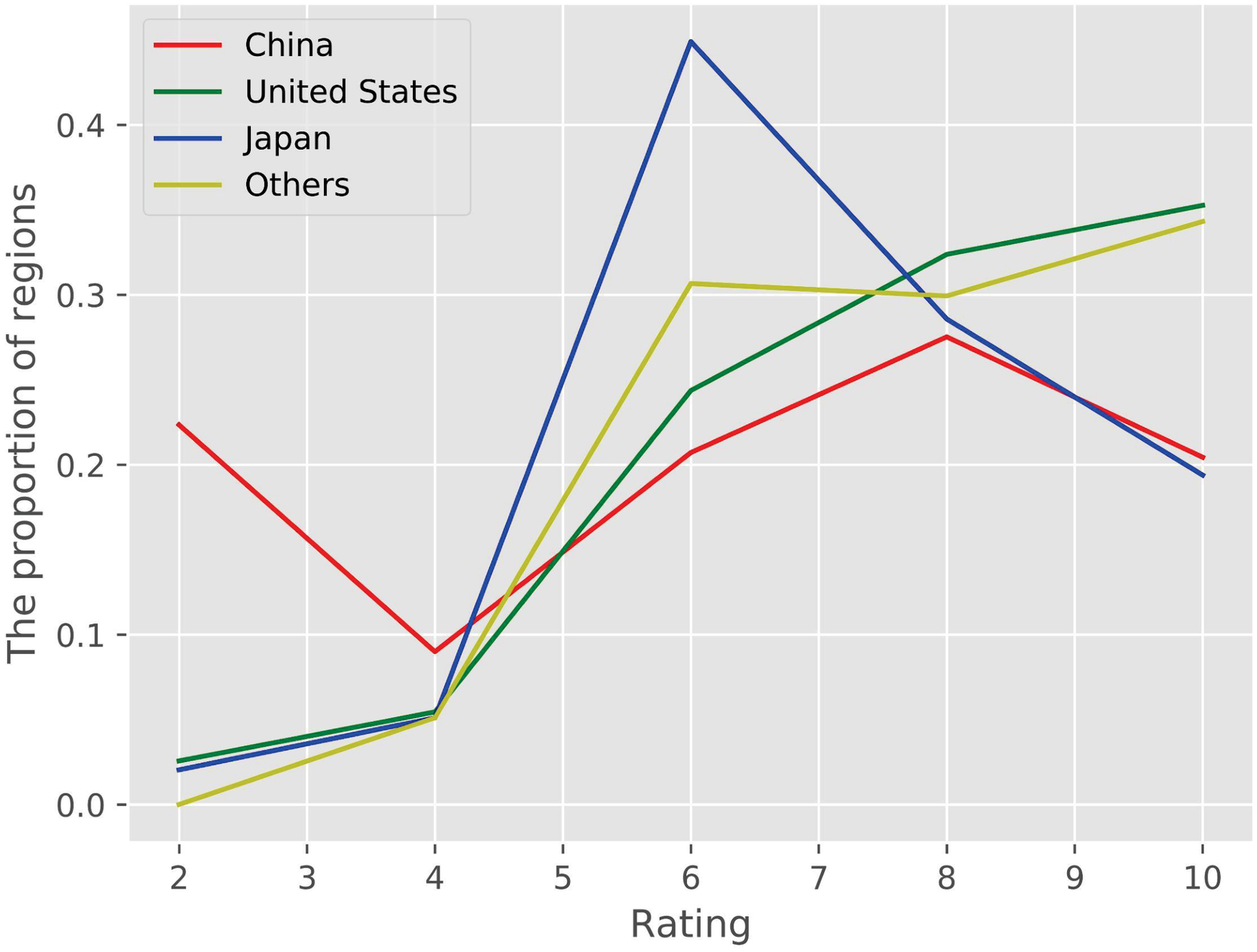}}
  \caption{Users' ratings for movies produced by different countries or regions.}
  \label{fig:region}
\end{figure*}
Similar to the users' attention to genres of movies, their rating distributions, which indicate users' expectations of movies in different regions, are also diverse and meaningful, as is shown in Fig.~\ref{fig:region}. Through combining movie elements and movie-related information to fuse the respective attributes into a joint model, a fine-grained user profile will be formed, and it could therefore facilitate user behavior modeling and review spam detection. 

\subsection{Sentence Embedding with Attention Mechanism}
Previous work generated sentence embeddings by composing word embeddings using operations on vectors and matrices. However, the simple average of the initial word embeddings does not work very well, since common words (e.g. 'the', 'and') have the same weight as key words in the sentence. The directed inspiration for our work is the smooth inverse frequency (SIF) method \cite{arora2016simple}, which computes the weighted average of the word vectors in the sentence and then removes the projections of the average vectors on their first singular vector. Especially, the word vectors are obtained from SkipGram, which is trained on the whole dataset $\mathbb{D}$.

\subsubsection{Feature Words of Movie Elements}
After the formal definition of a user's attention to a movie, the feature word list needs to be built. It is proved in \cite{hu2004mining} that words in reviews always converge when users comment on product features. The same conclusion could be drawn for movie reviews according to the statistical results on labeled data \cite{zhuang2006movie}, and a few words are able to capture most features, which are saved as the main part of the feature word list.
\begin{table}[b!]
\small
\renewcommand{\arraystretch}{1.3}
  \centering
  \caption{Basic feature word list of movie elements.}
    \begin{tabular}{c|c}
    \hline
        Element class&Feature words\\
        \hline
        \hline
        Plot      &story, plot, script, logic, scenario\\
        \hline
        Visual Effect &scene, picture, scenery, shot, editing\\
        \hline
        Sound Effect  &music, song, sound, soundtrack, theme\\
        \hline
        Acting Skill  &character, role, acting, actor, cast\\
        \hline
    \end{tabular}
\label{tab:words}
\end{table}

Table~\ref{tab:words} shows the basic feature words of movie elements. For a better generalization ability, five representative words for each aspect are selected in advance. Afterward, words with highest vector similarity to that in Table~\ref{tab:words} are picked out, and those with high frequency in movie review corpus are added into feature word list as well. If the above feature words appear in a user's reviews, they are perceived as his or her attention to the corresponding movie elements.

\subsubsection{Modified SIF Method}
The SIF method is based on the latent variable generative model \cite{arora2016latent} for text. The model regards corpus generation as a dynamic process, where the $t$-th word is produced at step $t$. The process is driven by the random sampling of a discourse vector $v_t \in \mathbb{R}^d$, which changes over time and represents the state of the sentence. The inner product between the discourse vector $v_t$ and the vector $v_w$ of word $w$ indicates the correlations between the word and the discourse. The probability of observing a word $w$ at time $t$ is calculated by a log-linear word production model:
\begin{equation}
Pr[w {\rm\ emitted\ at\ time}\ t\ |\ v_t]\propto exp(\langle v_t,v_w\rangle).
\end{equation}
All the $v_t$'s in the sentence $s$ are replaced for simplicity by a fixed discourse vector $v_s$, since the discourse vector $v_t$ does not change much \cite{arora2016simple}.

However, some words occur out of context, and some common words often appear regardless of the discourse. Two types of smoothing terms are therefore introduced in the log-linear model. The additional term $\alpha p(w)$ allows words to occur even if their vectors have low inner products with $v_s$, where $p(w)$ is the unigram probability of a word in the corpus. The common discourse vector $v_0 \in \mathbb{R}^d$ serves as a correction term for the most frequent discourse that is often related to the syntax, which increases the co-occurrence probability of words that have a high component along $v_0$. Noted that though feature words $w_f$ are frequent in the entire corpus, which lead to a very large $p(w)$, there is a strong correlation between $w_f$ and $v_s$. In order to improve the connection between sentence vectors and feature words, the value of scalar $\alpha$ is constrained so that users' attention could be better captured. Concretely, the probability of a word $w$ in the sentence $s$ is modeled by:
\begin{equation}
\left\{
\begin{array}{rl}
&Pr[w {\rm\ emitted\ in\ sentence}\ s\ |\ v_s]\\
&\qquad \qquad \qquad \qquad = \alpha p(w) + (1-\alpha)\frac{exp(\langle \widetilde v_s,v_w\rangle)}{Z_{\widetilde v_s}}, \\
&s.t.\ \ \alpha=0 {\rm\quad if\ } w {\rm\ in\ feature\ word} {\rm\ list},
\end{array}
\right.
\end{equation}
where $\widetilde v_s = \beta c_0 + (1-\beta)v_s$ and $v_0 \perp v_s$. $\alpha$ and $\beta$ are hyperparameters, and $Z_{\widetilde v_s} = \sum_{w\in V} exp(\langle v_t,v_w\rangle)$ is the normalizing constant. The equation allows a word $w$ that unrelated to the discourse $v_s$ to be emitted according the two terms: the unigram probability in the entire corpus of the word $w$, and the correlation with the common discourse vector $v_0$. The two terms are controlled by the scalar $\alpha$, which is set to 0.95 since $p(w)$ is several orders of magnitude smaller than the latter term. A wide range of the parameter $\beta$ can achieve close-to-best results, and the value is 0.1 in the experiment, which indicates that $v_s$ is more important in corpus generation. A larger $\beta$ will magnify the role of the syntax and cause the corpus to be homogeneous.

Feature words appear frequently in the corpus thus their $p(w)$ are large, which means they could be related to the syntax and unimportant. Therefore, the value of $\alpha$ is constrained to 0 for feature words, thus they could fit closely into the main idea of the sentence $v_s$. The final sentence embedding is defined as the maximum likelihood estimation for the vector $v_s$, which is the weighted average of the word embeddings in the sentence. By referring the derivation and solving process in \cite{arora2016simple}, the estimation for $\widetilde v_s$ is approximately,
\begin{equation}
\left\{
\begin{array}{rl}
&{\rm arg\,max} \sum\limits_{w\in s} f_w(\widetilde v_s) \propto \sum\limits_{w\in s}\frac{1-\alpha}{1+\alpha(Zp(w)-1)}v_w, \\
&s.t.\ \ \alpha=0 {\rm\quad if\ } w {\rm\ in\ feature\ word} {\rm\ list},
\end{array}
\right.
\end{equation}
where the partition function $Z$ is roughly the same for all $\widetilde v_s$ since $v_w$ are roughly uniformly dispersed on their assumption. Since $Z$ is fixed in all directions, the weight of word in the sentence depends only on the probability $p(w)$, and word that appears more frequently has smaller weight. It leads to a down sampling of the frequent words, whereas feature words are embedded in the sentence vectors without being down sampled, so that they are able to express the theme of reviews and catch users' attention. By calculating the first principal component of the vector $\widetilde v_s$, the direction $v_0$ could be estimated. The final sentence vector $v_s$ is obtained by subtracting the projection of $\widetilde v_s$ to their first singular vector $u$.
\begin{equation}
v_s=\widetilde v_s-uu^T\widetilde v_s.
\end{equation}
The latter term $uu^T\widetilde v_s$ serves as the common discourse vector in the whole corpus. The coupling degree of each sentence vector is reduced by removing the common component, which further enhance the robustness and make the theme of sentences clearer.

\subsection{Attetion-driven Conditional GANs}
Since there are a large number of normal reviews but few deceptive reviews, the model is supposed to learn the distribution of normal data from these samples. Based on the differences between the input test sample and normal reviews, the model could make a judgement about whether the sample is a spam or not. This approach is naturally corresponding to the idea of GANs. 

\begin{figure*}[htb!]
\centering
  \includegraphics[width=0.8\linewidth]{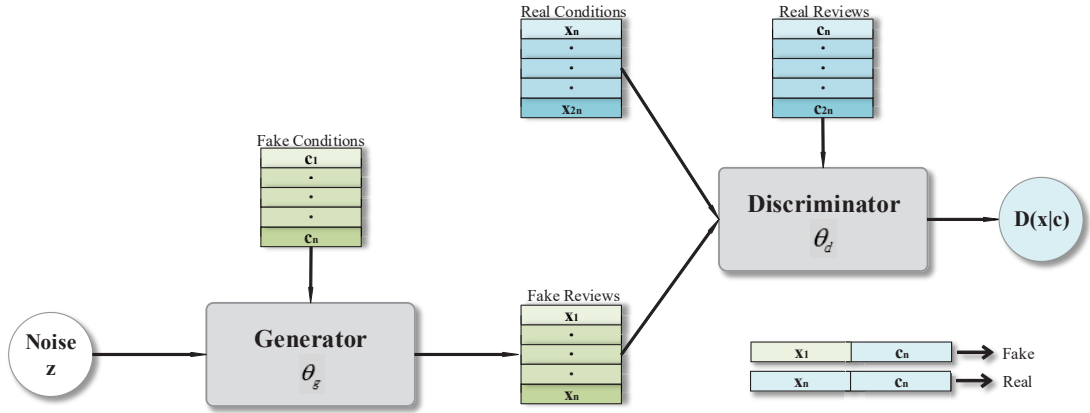}
\caption{Architecture of the modified conditional GAN. A batch of movie features are sampled and fed into the generator as conditions, in order to generate the reviews. Then another batch of movie features and the corresponding reviews are resampled from the training set and denoted as real. On the contrary, the resampled movie feature and generated review pairs are labeled fake. The discriminator is able to learn both linguistic features from reviews and the matching relation from these pairs, and therefore to identify deceptive reviews that does not match the linguistic style or the context.}
\label{fig:cgan}
\end{figure*}

Nevertheless, there are still two problems to be solved before GANs are utilized to review spam detection. Unlike traditional anomaly detection, each review corresponds to a particular movie, which means that deceptive reviews have no specific features or distribution. A review posted by this user is normal, while the identical one posted by another user could be a spam, and the same goes for different movies. On the other hand, if $\mathcal{D}$ converges well, it will fail to discriminate between real reviews and generated ones, which is contrary to our purpose. Therefore, the theoretical model is modified motivated by the above empirical observation. Although a single review cannot be judged to be truthful or deceptive, whether it follows the writing style that the user developed on similar movies before is a key entry point. The model draws on the idea of conditional GANs \cite{mirza2014conditional} and set the movie's additional information as conditions, and each user's data is trained into a separate model. Moreover, the architectural details of $\mathcal{G}$ and $\mathcal{D}$ are referred to \cite{arjovsky2017wasserstein}, considering the superiority of the Wasserstein GAN. Towards more efficient modeling, the conditional GAN changes as follows, whose architecture is shown in Fig.~\ref{fig:cgan}.

\textbf{Generator.} Given the joint distribution $P(v,c_g)$ of sentence vectors $v$ and conditions $c_g \in \mathbb{C}$ from real samples, $\mathcal{G}$ aims to learn a parameterized conditional distribution $P(v,z,c_g,\theta_g)$ that best approximates the true distribution. The prior input noise vectors $z$ and additional information $c_g$ are combined in joint hidden representation, and the adversarial training framework allows for considerable flexibility in how the hidden representation is composed. The generated sentence vector is conditioned on the network parameters $\theta_g$, noise vectors $z$ and generator conditions $c_g$. The loss for $\mathcal{G}$ is defined as:
\begin{equation}
\mathcal{L}_{\mathcal{G}}=-\mathbb{E}_{z\sim P_z(z)}\,[\mathcal{D}(\mathcal{G}\,(z\,|\,c_g)\,|\,c_g)],
\end{equation}
where the noise vector $z$ is drawn from a noise distribution $p_z(z)$, and $\mathcal{G}\,(z\,|\,c_g)$ are sentence vectors generated by noise $z$ under generator conditions $c_g$. The goal of the generator $\mathcal{G}$ is to generate as realistic reviews as possible. Therefore, it achieves the optimal performance if the discriminator $\mathcal{D}$ cannot distinguish real reviews from the generated ones, i.e. the output of discriminator $\mathcal{D}(\mathcal{G}\,(z\,|\,c_g)\,|\,c_g)$ is close to 1. In the process, $\mathcal{G}$ and $\mathcal{D}$ work under the same conditions $c_g$, which means that they only focus on the linguistic features of reviews.

\textbf{Discriminator.} The sentence vector $v$ and the condition $c_d$ are concatenated as the input of $\mathcal{D}$, where $v$ is identified whether it is real or fake by computing the probability $Pr(v|c_d,\theta_d)$. Since the ultimate goal is to detect whether a review is deceptive through $\mathcal{D}$, it is therefore necessary to avoid the problem of being unable to distinguish real reviews from generated ones after sufficient training. Different conditions for $\mathcal{G}$ and $\mathcal{D}$ are creatively deployed during their training, so that the model is able to learn the distribution of review vectors as well as the correlations between conditions and reviews concurrently, which is assumed to be critical in Section~\ref{para:3}. The loss for $\mathcal{D}$ is defined as:
\begin{equation}
\mathcal{L}_{\mathcal{D}}=\mathbb{E}_{z\sim P_z(z)}\,[\mathcal{D}(\mathcal{G}\,(z\,|\,c_g)\,|\,c_d)]-\mathbb{E}_{x\sim P_{data}(x)}\,[\mathcal{D}(x\,|\,c_d)].
\end{equation}
On the condition of $c_d$, $\mathcal{D}$ is supposed to identify the matching reviews $x$ as real, and the generated reviews $\mathcal{G}\,(z\,|\,c_g)$ that match $c_g$ as fake. In other word, for the discriminator $\mathcal{D}$, the best outcome is to predict $\mathcal{D}(x\,|\,c_d)$ to be 1 and $\mathcal{D}(\mathcal{G}\,(z\,|\,c_g)\,|\,c_d)$ to be 0. Fake reviews vectors are produced under generator conditions $c_g$, while they are concatenated to discriminator conditions $c_d$ and labeled fake. Noted that real review vectors $x$ are matched to $c_d$, which is resampled from the dataset and completely different from $c_g$. As mentioned above, generated reviews are real under the condition of $c_g$ while they are fake under the condition of $c_d$, as users' attention varies with different genres of movies, corresponding to the fact that a spammer usually cannot evaluate the movie from the perspective of normal reviews. In this way, $\mathcal{D}$ could identify a review to be fake or deceptive if it does not match the context or the style of historical reviews.

\section{Experiments}
\label{para:4}
In this section, the proposed model is compared with six state-of-the-art anomaly detection methods in terms of precision, recall and F1-measure. Before evaluating the performance of the proposed approach, where and how to collect the dataset are discussed.

\subsection{Dataset}
\label{datasec}
In this paper, all reviews are collected from Douban, a huge Chinese online community where users share their reviews to express their feelings about movies and score the movies with one star to five stars. Although the approach and the comparison algorithms are unsupervised, reviews need to be manually labeled in order to verify the effectiveness of the models. Therefore, several ways to identify spammed movies and reviews are explored as follows.

\begin{figure*}[htb!]
\centering
  \includegraphics[width=0.6\linewidth]{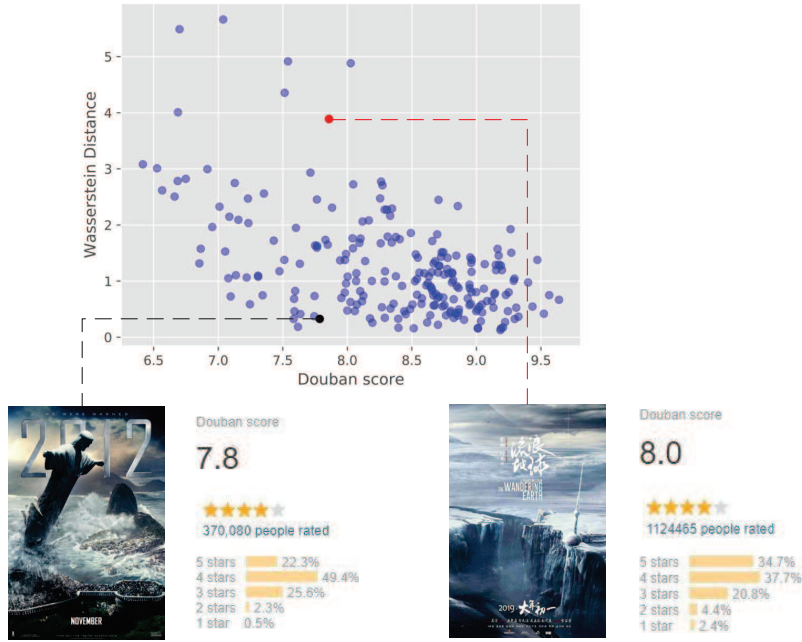}
\caption{The distribution of the Wasserstein distance between ratings of movies and the mean value, where the black dot \textit{2012} and the red dot \textit{The Wandering Earth} are highlighted.}
\label{fig:wd}
\end{figure*}

Due to the fact that spammers tend to give extreme ratings when they praise or criticize a movie driven by benifits, it is assumed that the distribution of ratings of normal movies is different from that of spammed movies, whose tendency toward polarization is more noticeable. Hence the Wasserstein metric is leveraged to find outliers in the distribution of ratings, which are suspected of being deceptive. Wasserstein metric is a distance function defined between probability distributions, and the distance between similar distributions is smaller. The Wasserstein distance between each movie and the datum reference is calculated, and the datum reference is the mean of ratings of all movies with same Douban score as the current one. The results are reported in Fig.~\ref{fig:wd}.

\begin{figure}[htb!]
\centering
  \includegraphics[width=0.8\linewidth]{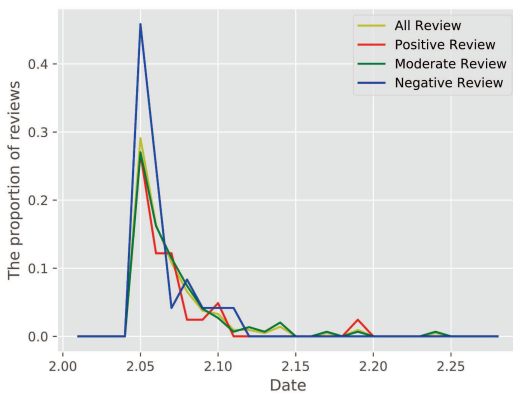}
\caption{The proportion of positive reviews, moderate reviews and negative reviews of \textit{The Wandering Earth} over time.}
\label{fig:time}
\end{figure}

It is evident that \textit{The Wandering Earth} is much larger than \textit{2012} on the proportion of one-star ratings and five-stars ratings, though their Douban score and genres of movies are similar. The word-of-mouth of \textit{The Wandering Earth} is polarised, which leads to a larger Wasserstein distance, hence it is more suspected of being deceptive. Moreover, the proportions of reviews with different ratings over time are investigated, as shown in Fig.~\ref{fig:time}.

It can be seen that the distributions of positive reviews and moderate reviews are roughly in line with that of all reviews, but most negative reviews are concentrated on February 5th, when \textit{The Wandering Earth} was released. It is also consistent with the analysis of the features of spammers that they will influence users' perception of movies by posting biased reviews when the movie is released. Combining the Wasserstein distance, the proportion of ratings and users' view, six spammed movies are identified, including \textit{The Wandering Earth}.

Douban displays 500 positive reviews, 500 moderate reviews and 500 negative reviews for each movie as top reviews. In determining specifically whether a review is deceptive, we solicit the help of five volunteer graduate students to make independent judgements on the dataset. The criteria for review spam come from \textit{30 Ways You Can Spot Fake Online Reviews} \footnote{https://consumerist.com/2010/04/14/how-you-spot-fake-online-reviews}, which provides some techniques to spot deceptive online reviews. Besides, the features of review spam presented in Section~\ref{para:3} are also utilized as a reference. In addition, The user review weighting system of Douban platform offers great help. In order to avoid the spammers' manipulation in the word-of-mouth of movies, Douban platform developed a weighting system for users' reviews and ratings according to the daily behaviors of users. Therefore, the help votes (how many people think that review is helpful) and the ranking in the review list are also considered as the judgement bases of review spam. The ranking of reviews by the help votes and the ranking of Douban are compared, and the differences in position in sorting are viewed as the suspicion level. More specifically, a review is more likely to be deceptive if it has numerous help votes but ranks low on Douban. The paper collects 13,056 reviews from 3,261 users on nine movies as the test set, in which 10,268 are truthful reviews and 2,788 are deceptive, then the rest of historical reviews of these users are used for training.

\begin{figure*}[htb!]
\centering
  \includegraphics[width=0.8\linewidth]{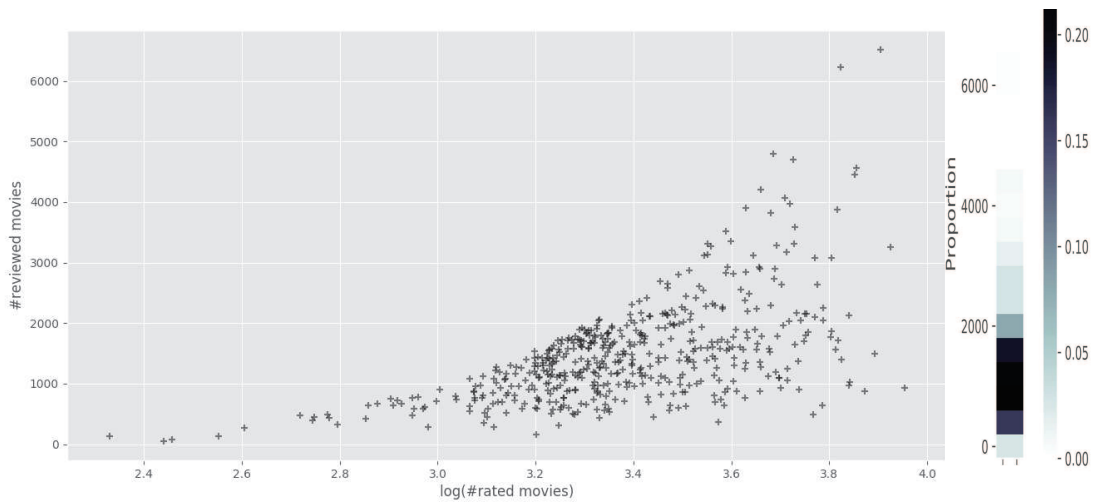}
\caption{The visualization of the number of reviewed movies and the log of rated movies for users.}
\label{fig:scatter}
\end{figure*}

Fig.~\ref{fig:scatter} shows that most users in the dataset posted more than 500 movie reviews, which facilitate us to mine their viewing preferences and review styles. Besides, through the meticulous investigation and analysis of users' interest in watching movies, six critical features are defined as conditions for generating user reviews, as illustrated in Table~\ref{tab:conditions}. These features are encoded in different ways and concatenated to embedding vectors of review, and the dimension of that is set 100. The average director score means the average of user's ratings for previous movies of this director that user has rated or commented, and the average actor score is also calculated in this way. 
\begin{table}[!t]
\small
\renewcommand{\arraystretch}{1.3}
  \centering
  \caption{The influencing factor of the emotional tendency of user reviews.}
    \begin{tabular}{c|c|c}
    \hline
        Conditions&Encoding mode&Dimension\\
        \hline
        \hline
        Movie score             &Real-Number Encoding &1\\
        \hline
        User rating             &Real-Number Encoding &1\\
        \hline
        Region                  &One-Hot Encoding     &4\\
        \hline
        Movie genres            &One-Hot Encoding     &20\\
        \hline
        Average director score  &Real-Number Encoding &1\\
        \hline
        Average actor score     &Real-Number Encoding &1\\
        \hline
    \end{tabular}
\label{tab:conditions}
\end{table}

\subsection{Baselines and Parameter Settings}
Since the paper is the first to do research on the movie review spam detection and innovatively propose the attention mechanism of reviewers, it is impractical and unfair to directly compare with other review spam detection algorithms that focus on the product. Moreover, all of these review spam detection algorithms are supervised or semi-supervised, to the best of our knowledge. Therefore, the condition vectors mentioned in Section~\ref{datasec} are concatenated with the review vectors together and six state-of-the-art unsupervised outlier detection algorithms are evaluated on the dataset.

Local Outlier Factor (LOF) \cite{breunig2000lof} is related to density-based clustering. The outlier factor is local in the sense that only a restricted neighborhood of each object is taken into account, and the degree depends on how isolated the object is concerning the surrounding neighborhood. The single parameter of LOF is the number of nearest neighbors used in defining the local neighborhood of the object, and 10 is the best tuning.

One-class SVM (OCSVM) \cite{scholkopf2001estimating} estimates a function $f$ to learn the underlying probability distribution of the dataset. The functional form of $f$ is derived by a kernel expansion in terms of a subset of the training data and regularized by controlling the length of the weight vector in an associated feature space. The Gaussian kernel is utilized in the experiment.

Multi-class SVM (MCSVM) \cite{liu2005one} \cite{quinn2019british} is a non-probabilistic binary linear classifier. It follows a similar setting with OCSVM. As multi-class SVM cannot handle the case that the labels of all training samples are consistent, unmatched movie-review pairs, which are introduced in the training of the discriminator $\mathcal{D}$, serve as the fake samples.

Minimum Covariance Determinant (MCD) \cite{hardin2004outlier} is a distributional fit to Mahalanobis distances which uses a robust shape and location estimate. The MCD of data points is the mean and covariance matrix based on the sample of a size that minimizes the determinant of the matrix. The contamination, which controls the proportion of abnormal samples, is 0 in the dataset.

Isolated Forest (iForest) \cite{liu2008isolation} builds an ensemble of isolated trees for a given dataset, then anomalies are those instances which have short average path lengths on the isolated trees. It detects anomalies based on the concept of isolation without employing any distance or density measure. There are two variables in this method: the number of trees to build and the sub-sampling size, and they are set to 100 and 256 respectively depending on the size of the dataset.

Variational Autoencoder (VAE) \cite{an2015variational} uses reconstruction probability in anomaly detection. It utilizes the generative characteristics of the variational autoencoder to derive the reconstruction of the data, thus it could analyze the underlying cause of the anomaly. The network structure of the encoder and decoder is set similarly to that of the generator and discriminator.

\subsection{Complexity Analysis and Efficiency Evaluation}
In the proposed algorithm, the preprocessing of reviews, i.e. the sentence embedding, for both training and testing phase consist of two part, respectively corresponding to the SkipGram model and the SIF method. Specifically, the computational complexity of SkipGram is $\mathcal{O}(c\gamma l\,{\rm log}n)$, where $c$ is the window size, $n$ is the total number of words, $\gamma$ is the number of sampling and $l$ is the length of the sampling sequence. For the SIF method, its complexity is $\mathcal{O}(msd)$, in which $s$ is the number of sentences, $m$ is the average number of words in each sentence, and $d$ is the embedding dimension. The total number of words $n$ is larger than the product of $m$ and $s$, since there will be repeated words in sentences.

For the conditional GAN, the computational complexity can be represented as $\mathcal{O}(DET/B)$, in which $D$ is the number of training samples, $E$ represents epochs, $B$ is the batch size, and $T$ is the computational complexity in each iteration. The generator and discriminator are composed of fully connected layers, dropout, batch normalization and the activation functions, with fully connected layers taking up most of the operation time. Therefore, $T$ can be further expressed as $\mathcal{O}(T)\approx\mathcal{O}(\sum_{i=2}^{L}{N^iN^{i-1}})$, in which $N^i$ is the number of neurons in the $i$-th layer, and $L$ is the number of layers in the network.

The computational complexity of neural network is hard to accurately describe due to many free variables, so the number of floating-point operations (FLOPs) is introduced to show the complexity of the model intuitively. The FLOPs of fully connected layers can be computed as:
\begin{equation}
{\rm FLOPs} = (2I-1)O,
\end{equation}
where $I$ is the input dimensionality and $O$ is the output dimensionality. Similar to the computational complexity, the time consumption of batch normalization and activation functions are not taken into account, since they are negligible compared with fully connected layers. The number of parameters and FLOPs are reported in Table~\ref{tab:para}.

\begin{table}[!t]
\small
\renewcommand{\arraystretch}{1.3}
  \centering
  \caption{Model Complexity of the Condition GAN.}
    \begin{tabular}{c|c|c|c}
    \hline
    \bf{Module}&\bf{\quad\ \ Layer \quad\qquad}&\bf{ Parameters }&\bf{\quad FLOPs\quad\quad}\\
    \hline
    \hline
        \multirow{5}{*}{\shortstack{\bf{Generator}\\ \\ \bf{$\mathcal{G}$}}}&fc-1&33,024&65,536\\
        \cline{2-4}
        &fc-2&131,584&262,144\\
        \cline{2-4}
        &fc-3&131,328&262,144\\
        \cline{2-4}
        &output&25,700&51,200\\
        \cline{2-4}
        &\bf{total}&\bf{321,636}&\bf{641,024}\\
        \hline
        \hline
        \multirow{5}{*}{\shortstack{\bf{Discriminator}\\ \\ \bf{$\mathcal{D}$}}}&fc-1&33,024&65,536\\
        \cline{2-4}
        &fc-2&32,896&65,536\\
        \cline{2-4}
        &fc-3&8,256&16,384\\
        \cline{2-4}
        &output&65&128\\
        \cline{2-4}
        &\bf{total}&\bf{74,241}&\bf{147,584}\\
        \hline
    \end{tabular}
\label{tab:para}
\end{table}

There are 395,877 trainable parameters for the conditional GAN, and the total number of FLOPs of the model is 788,608. The computational efficiency is evaluated on a NVIDIA Tesla V100 GPU. In the training phase, the mini-batch SGD \cite{li2014efficient} is employed as the optimizer, and the batch size is set to 128. It takes less than 15ms for the training of a mini batch of samples in each iteration, including the forward and backward propagation. In the testing phase, the inference process for each review can be completed in 0.1ms, as it only requires one forward propagation of the discriminator, which ensures a real-time online review spam detection.

\begin{figure}[!htbp]
\centering
  \includegraphics[width=0.75\linewidth]{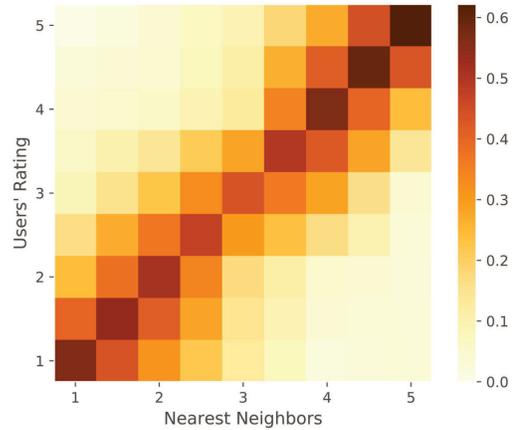}
\caption{The rating distribution of the nearest neighbors of reviews with a particular rating.}
\label{fig:heat}
\end{figure}

\subsection{Results}
Before comparing with other algorithms, the effectiveness of the proposed method is verified. Vectors mapped by reviews utilizing the modified SIF method are aggregated, and the nearest neighbor approach is used to find out whether a review has the same rating as its neighbors. For reviews with the same rating, multiple tests are conducted to ensure the accuracy of the experiment, and the results are represented by the proportions of neighbors with different ratings.

\begin{figure*}[b!]
  \centering
  \subfigure[The loss of $\mathcal{G}$ and $\mathcal{D}$.]{
    \includegraphics[width=0.4\linewidth]{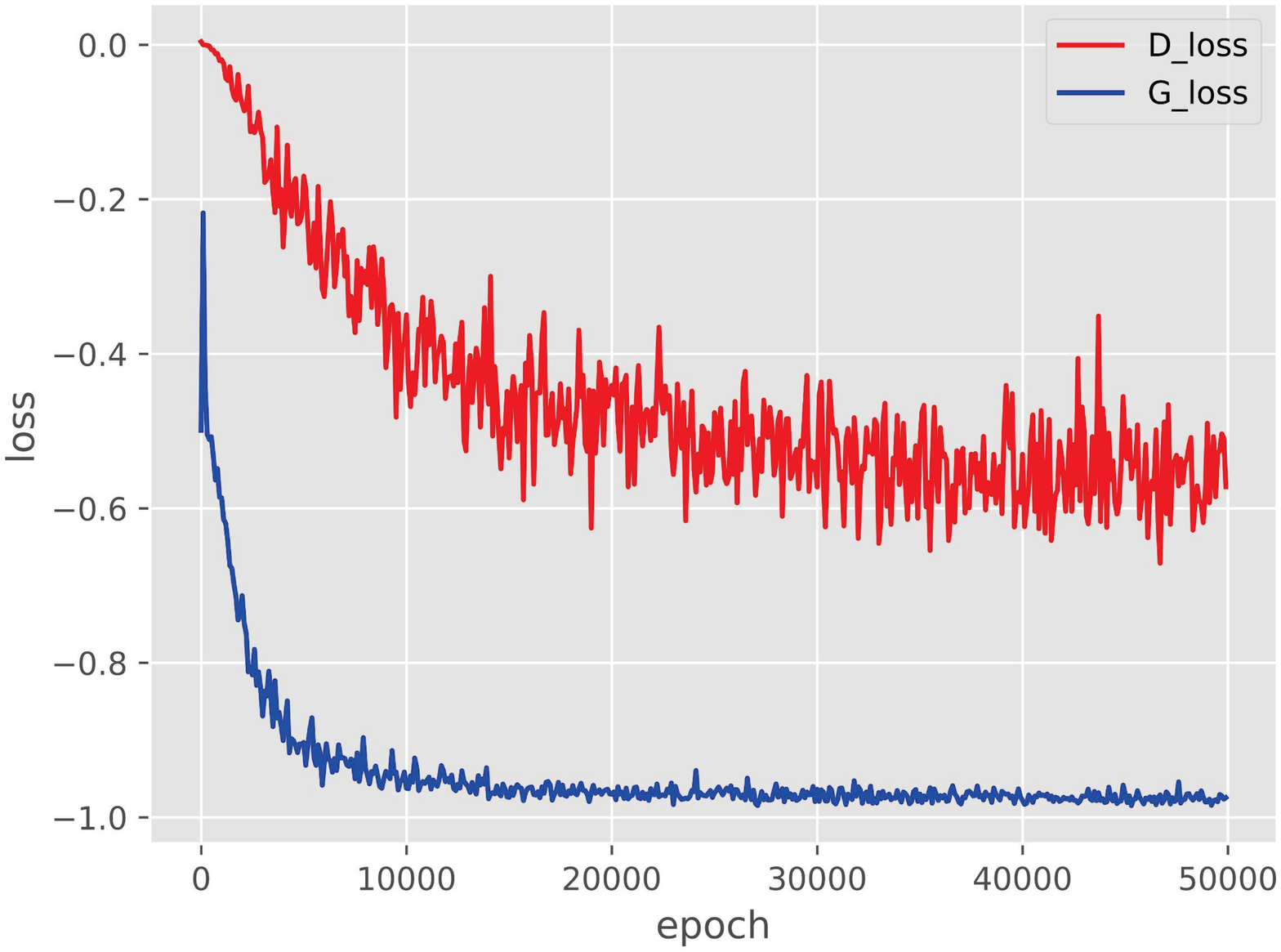}
    \label{fig:la}}
  \subfigure[The discriminant probability of $\mathcal{D}$]{
    \includegraphics[width=0.393\linewidth]{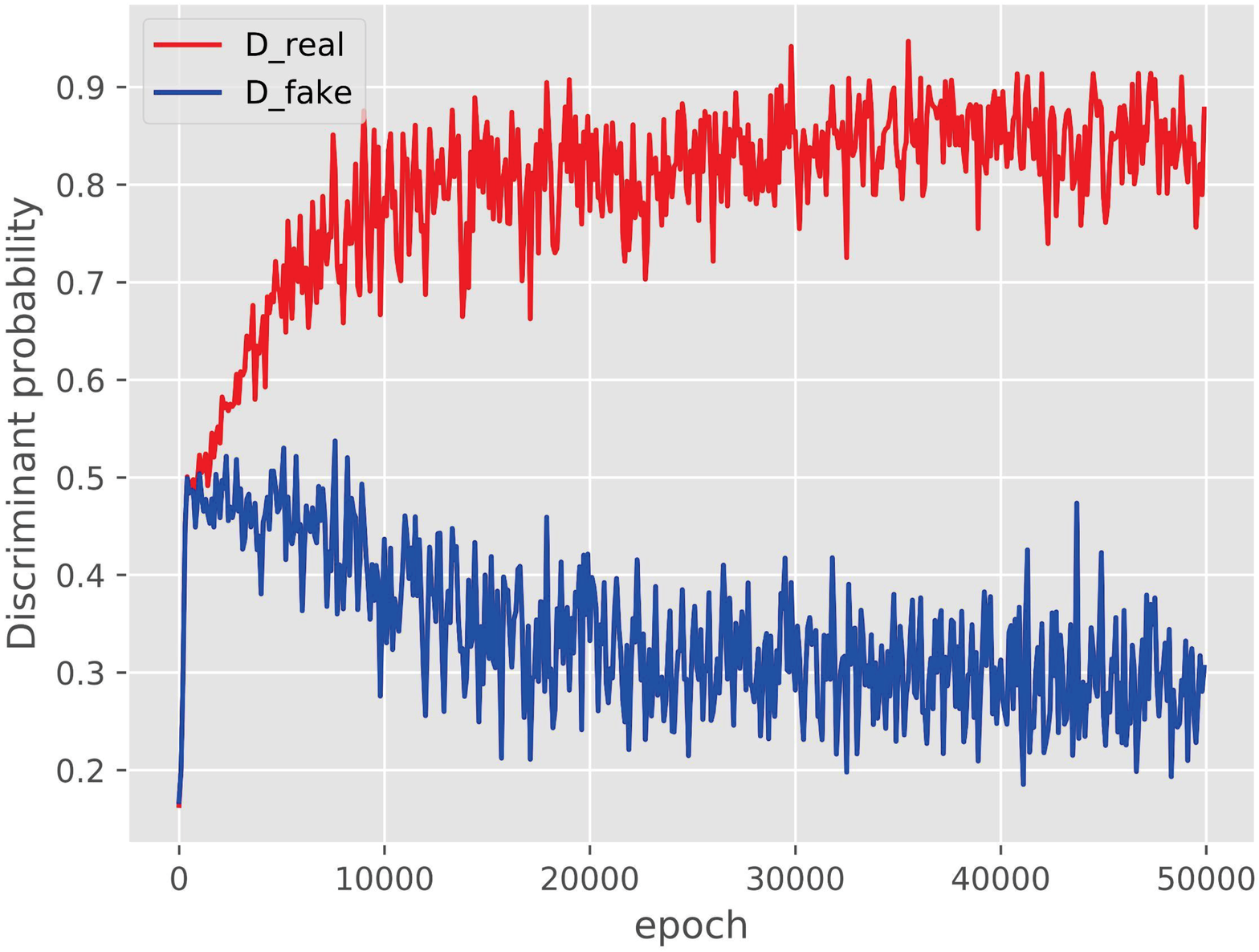}
    \label{fig:lb}}
  \caption{The loss and discriminant probability of the conditional GAN.}
  \label{fig:loss}
\end{figure*}

The secondary diagonal of the similarity matrix in Fig.~\ref{fig:heat} represents an accurate clustering in the five-star rating system, and it indicates that the embedding method is able to maintain the original attributes and meaning of the positive review and negative review. Because of the strong subjectivity of users' ratings and reviews, it does not perform well on clustering moderate reviews, but it can still roughly assess a reviewer's attitude toward the movie. On this basis, the additional information and review vectors are fed into the conditional GAN for training and learning. Noted that the loss functions of $\mathcal{G}$ and $\mathcal{D}$ are not following those of the original conditional GAN. They work together to identify fake reviews instead of competing with each other, and the loss curve is shown in Fig.~\ref{fig:loss}.

$\mathcal{G}$ aims to fool $\mathcal{D}$ and let it judge the generated vectors to be true, and the loss of $\mathcal{G}$ reaches the minimum value at around 10,000 epochs, which means generated vectors can completely deceive $\mathcal{D}$. The goal of $\mathcal{D}$ is to distinguish between real reviews that match conditions and generated reviews that do not. For example, a one-star review which is full of praise is undoubtedly a fake review. Since conditions are randomly assigned to generated vectors, it is inevitable that random condition is similar to the real condition, especially for users who have single viewing preference and rating habit. Therefore, $\mathcal{D}$ is better at identifying real samples than fake ones, as illustrated in Fig.~\ref{fig:lb}.

\begin{table*}[thbp]
\small
\renewcommand{\arraystretch}{1.3}
  \centering
  \caption{Comparison with baselines.}
    \begin{tabular}{c|c|c|c|c|c|c|c}
    \hline
        \multirow{2}{*}{\bf{Approach}}&\multirow{2}{*}{\bf{Accuracy}}&\multicolumn{3}{c|}{TRUTHFUL}&\multicolumn{3}{c}{DECEPTIVE}\\
        \cline{3-8}
        &&\bf{Precision}&\bf{Recall}&\bf{F1-score}&\bf{Precision}&\bf{Recall}&\bf{F1-score}\\
        \hline
        \hline
        LOF           &79.66&89.99&83.42&86.58&51.88&65.82&58.02\\
        \hline
        OCSVM         &79.96&81.16&\bf{97.06}&88.40&61.08&17.00&26.60\\
        \hline
        MCSVM         &80.96&83.08&95.17&88.72&61.67&28.62&39.10\\
        \hline
        MCD           &81.19&84.54&93.11&88.62&59.51&37.27&45.83\\
        \hline
        iForest       &82.47&85.44&93.67&89.37&63.87&41.21&50.10\\
        \hline
        VAE  		  &83.93&87.14&93.34&90.13&66.76&49.28&56.71\\
        \hline adCGAN\underline{~~}w/o\underline{~~}attention
         			  &84.20&88.37&92.02&90.16&65.34&55.38&59.95\\
        \hline
        adCGAN (ours) &\bf{87.03}&\bf{92.97}&90.33&\bf{91.63}&\bf{67.76}&\bf{74.86}&\bf{71.13}\\
        \hline
    \end{tabular}
\label{tab:result}
\end{table*}
Afterward, the performance of the algorithm is evaluated on the dataset crawled from Douban. Several standard evaluation metrics are adopted: accuracy, precision, recall and F1-score, which are computed using a micro-average, i.e., from the aggregate true positive, false positive and false negative rates, as suggested in \cite{forman2010apples}. The above evaluation metrics of truthful samples and deceptive samples are computed respectively, and the comparison results between our model and state-of-the-art methods are shown in Table~\ref{tab:result}.

It is clear that the proposed adCGAN achieves the best performance in terms of accuracy, precision and F1-score on both truthful and deceptive reviews. Since the training samples may contain some deceptive reviews, linear models, which map the data and learn their boundaries (e.g. OCSVM, MCSVM and MCD), have poor performance on review spam detection. Note that OCSVM achieves the highest recall rate for truthful reviews but the lowest recall rate for deceptive ones, which means that support vector machines indiscriminately identify a majority of the testing samples as truthful due to the wrong boundaries. Leveraging unmatched movie-review pairs, MCSVM performs a little better than OCSVM, but it still faces the problem of identifying most of the deceptive reviews as truthful ones. For iForest, a random dimension is selected to build a tree, but a large amount of dimension information is not utilized, which are not mutually independent in review vectors, resulting in reduced reliability of the algorithm. To our surprise, another simple density-based model, LOF, is able to detect a majority of deceptive reviews, though it comes at the cost of minimal accuracy. The VAE architecture detects outliers by calculating reconstruction probability. The difference between VAE and adCGAN is that many generated samples with fake conditions are leveraged in the presented approach in the process of judgement (reconstruction), which is equivalent to demarcate the data in the original space and conducive to learning their distribution. Moreover, the result of the model without attention mechanism (adCGAN\underline{~~}w/o\underline{~~}attention) demonstrates the assumptions in Section~\ref{para:3} that a normal user usually has relatively consistent attention towards specific movies. The presented adCGAN can not only generate highly effective vectors whose distribution is close to that of real review vectors, but also determine if a movie review is derived from the corresponding additional information, and hence it achieves significantly better performance than baseline algorithms.

\section{Discussion, Implication, and Conclusion}
\label{para:5}
In this paper, a novel adCGAN model is developed for movie review spam detection. To address the problem of lacking labeled training samples, a new implementation of GAN is introduced by resampling conditions from the dataset. It also leverages the attention mechanism to facilitate GAN to capture the correlations of movie-review pairs, which improves the accuracy of the model. The experimental results on Douban dataset demonstrate the superior performance on movie review spam detection. 

The idea of unmatched training pairs and the attention mechanism can be migrated to various review spam detection, such as literature reviews and other types of reviews that reflect users’ consistent interests. However, it may not be well suited to product reviews, since they are more related to the quality of product, while the adCGAN model is user-centric and focuses on user preference. Furthermore, it is necessary for the proposed method to set the feature words manually, thus element class and feature words need to be reinvestigated for other types of reviews, which limits the generality of the model.

The following directions can be explored in the future:

(1) The paper has investigated the effectiveness of adCGAN on movie review spam detection. The generalization ability of the model can be further improved by mining the inherent attributes of reviews, in order to make it scalable for other types of review spam detection. 

(2) The adCGAN model could simultaneously capture review styles and viewing preferences, which are embedded in the weight of GAN. It is attractive to extract the rich information in the follow-up work and leverage it in recommendation systems.

\ifCLASSOPTIONcaptionsoff
\newpage
\fi



%\bibliographystyle{IEEEtran}

\bibliography{ref}



%\vfill


\end{document}